\documentclass[preprint, aps,nofootinbib,tightenlines, byrevtex]{revtex4}
\usepackage{graphicx}%
\usepackage{dcolumn}
\usepackage{amsmath}
\usepackage{epsfig}
\usepackage{latexsym}
\usepackage{amssymb}
\usepackage{mathrsfs}

\newcommand{\3}{\ss}

\newcommand{\MeV}{\ensuremath{\mathrm{MeV}}}
\newcommand{\fm}{\ensuremath{\mathrm{fm}}}

 \newcommand{\calD}{\mathcal{D}}
\newcommand{\calH}{\mathcal{H}} \newcommand{\calK}{\mathcal{K}}

\newcommand{\dis}{\displaystyle}
\newcommand{\ii}{\mathrm{i}}
\newcommand{\dd}{\mathrm{d}}

\begin{document}
\bibliographystyle{plain}

\title{Effective Field Theory Calculation of $nd$ Radiative Capture at Thermal
  Energies}

\author{H. Sadeghi$^{a}$}\email{H-Sadeghi@araku.ac.ir}
\author{S. Bayegan$^{b}$}\email{Bayegan@Khayam.ut.ac.ir}
\author{Harald W. Grie{\ss}hammer$^{c}$}\email{hgrie@gwu.edu}
 \affiliation{$^a$ Department of Physics, University of Arak, P.O.Box 38156-879, Arak, Iran.\\
 $^b$Department of Physics, University of Tehran, P.O.Box 14395-547, Tehran, Iran.\\
 $^c$Center for Nuclear Studies, Department of Physics,\\
The George Washington University, Washington DC 20052, USA.}

\vspace{2cm}

\begin{abstract}

\vspace{0cm}
 The cross section for the thermal neutron capture by the deuteron is
 calculated with pionless Effective Field Theory(EFT). No new Three-Nucleon
  forces are needed up to next-to-next-to-leading order in order to achieve
cut-off independent results, besides those fixed by the triton
binding energy and Nd scattering length in the triton channel. The
cross-section is accurately determined to be $\sigma_{tot}=[0.503\pm
0.003]mb$. At zero energies, the magnetic $M1$-transition gives the
dominant contribution and is calculated up to
next-to-next-to-leading order (N$^2$LO). Close agreement between the
available experimental data and the calculated cross section is
reached. We
demonstrate convergence
and cutoff independence order by order in the low-energy expansion.\\

\begin{tabular}{c}
\end{tabular}

\begin{tabular}{rl}

keywords:&\begin{minipage}[t]{11cm} effective field theory,
three-body system, three-body force, Faddeev equation, radiative
capture
  \end{minipage}

\end{tabular}
\end{abstract}

\vskip 1.0cm \noindent

\maketitle

\section{Introduction}
  \label{introduction}

The study of the three-body nuclear system involving neutron
radiative capture by deuteron has been investigated in theoretical
and experimental works over the past years. The experimental result
of this process  has most accurately been measured by Jurney, et.al.~\cite{Jurney}.
The value of 0.508$\pm$0.015(mb) for the cross section was resulted
for 2200 m/sec neutrons.

Rapid  progress has been made in the theoretical study of the
$Nd\rightarrow{^3H}\gamma $ reaction such as  the $p$-$d$ and
$n$-$d$ radiative capture.  At such energies a magnetic
dipole($M_1$)transition is almost entirely participated. These
reactions were studied in plane wave (Born) approximation by Friar
et al.~\cite{friar}. In these investigations the authors employed
their configuration-space Faddeev calculations of the helium wave
function, with inclusion of three-body forces and pion exchange
currents.  More recently a rather detailed investigation of such
processes has been performed by Viviani et~al.~\cite{Viviani,19}. In
their calculations  the quite accurate three-nucleon bound- and
continuum states were obtained in the variational pair-correlated
hyperspherical method from a realistic Hamiltonian model with two-
nucleon and three- nucleon interactions.

They obtained in Ref.~\cite{Viviani} the cross section from Argonne $v_{14}$
two-nucleon and Urbana VIII three-nucleon interactions(AV14/UVIII), also from
Argonne $v_{18}$ two-nucleon and Urbana IX three-nucleon
interactions(AV18/UIX) and including $\Delta$ admixtures. Cross section values
were found 0.600 (mb) and 0.578 (mb) which overestimate the experimental value
by 18$\%$ and 14$\%$ value, respectively, see table 2. It should be noted,
however, that the explicit, non-perturbative inclusion of $\Delta$-isobar
degrees of freedom in the nuclear wave function are found to be in
significantly better agreement with experiment than those obtained from
perturbative ($\Delta_{PT}$) estimates. This shows that these results for this
very-low energy observable are sensitive to details of the short-range part of
the interaction.  recent calculation using manifestly gauge-invariant currents
reduced the spread~\cite{19}, but the result including three-body currents,
$0.558$ mb, still over-predicts the cross-section by 10\%. Model-dependent
currents associated with the $\Delta(1232)$ were identified as source of the
discrepancy. Thus, the question remains how such details of short-range
Physics can so severely influence a very-long-range reaction with maximal
energies of less than $10$ MeV.

During the last few years, nuclear Effective Field Theory(EFT) has been
applied to two-, three-, and four-nucleon systems, see
e.g.~\cite{kaplan,Beane, 3stooges_boson,3stooges_doublet,
  doubletNLO,pbhg,chickenpaper}. The pionless Effective Field Theory would be
an ideal tool to calculate low-energy cross sections in a model-independent
way and to possibly reduce the theoretical errors by a systematic,
model-independent calculation with an a-priori estimate of the theoretical
uncertainties.  An example of a precise calculation is the reaction
$np\rightarrow \gamma d$, which is relevant to big-bang nucleosynthesis(BBN).
The cross section for this process was computed to $1\%$ error for center of
mass energies $E\lesssim 1 Mev$~\cite{Rupak98,Rupak99,Ando06}.

 We  have suggested a method for
computation of neutron-deuteron radiative capture for extremely
low energy( $ 20 \leq E \leq 200 $ Kev )with pionless
EFT~\cite{Sadeghi}, where with this formalism, we can estimate
errors in a perturbative expansion up to N$^2$LO within a few percent of the
ENDF values~\cite{ENDF}.

The purpose of the present paper is to study the cross section for
radiative capture of neutrons by deuterons $nd\rightarrow \gamma{^3H}$ at zero
energies with pionless EFT. At these energies, the magnetic $M_1$-transition
gives the dominant contribution. The $M_1$ amplitude is calculated up to
next-to-next-to-leading order (N$^2$LO) with insertion of three body force.
Results show less than $1$\% deviation from the available experimental data at
zero energy ($0.0253$ eV).

This article is organized as follows. In the next section, a brief description
of the formalism and its input for total cross section of the neutron-deuteron
radiative capture will be presented. We discuss the theoretical errors,
tabulation of the calculated cross section in comparison with the other
theoretical approaches and the most recent data~\cite{Jurney}
in section~\ref{section:results}.  Finally, Summary and conclusions follow in
Section~\ref{section:conclusion}.

\section{ Neutron-deuteron scattering in triton channel and radiative capture}
\setcounter{equation}{0} \label{section:Formalisem}

The $^2\mathrm{S}_{\frac{1}{2}}$ channel to which $^3$He and $^3$H belong is
qualitatively different from the other three-nucleon channels because all
three nucleons can occupy the same points in space. Consequently, $^2S_{1/2}$
describes the preferred mode for $nd\rightarrow^3H\gamma$ and
$pd\rightarrow^3He\gamma$.  The three-nucleon Lagrangean is well-known and
will not be repeated here, see e.g.~[14,18] for details.

The derivation of the integral equation describing neutron-deuteron
scattering has also been discussed
before, see e.g.~\cite{3stooges_doublet,griesshammer}. We present here only
the results. The integral equation is solved numerically by imposing
a cut-off $\Lambda$. In that case, a unique solution exists in the
$^2S_{1/2}$-channel for each $\Lambda$ and vanishing three-body force, but no unique
limit as $\Lambda\to\infty$.  As long-distance phenomena must
however be insensitive to details of the short-distance physics (and
in particular of the regulator chosen), Bedaque et
al.~\cite{3stooges_boson,3stooges_doublet,4stooges,griesshammer}
showed that the system must be stabilized by a three-body force
\begin{equation}
  \label{eq:calH}
   \calH(E;\Lambda)=
   \frac{2}{\Lambda^2}\sum\limits_{n=0}^\infty\;H_{2n}(\Lambda)\;
   \left(\frac{ME+\gamma_t^2}{\Lambda^2}\right)^n
   =\frac{2H_0(\Lambda)}{\Lambda^2}+
   \frac{2H_2(\Lambda)}{\Lambda^4}\;(ME+\gamma_t^2)+\dots \;.
\end{equation}
which absorbs all dependence on the cut-off as $\Lambda\to\infty$.
It is analytical in $E$ and can be obtained from a three-body
Lagrangean, employing a three-nucleon auxiliary field analogous to
the treatment of the two-nucleon channels~\cite{4stooges}. Contrary
to the terms without derivatives, there are different, inequivalent
three-body force terms with \emph{two} derivatives, but only one of
them, $H_2$, is enhanced over its naive dimensional estimate,
mandating its inclusion at N$^2$LO~[14,20]. Neutron-deuteron
scattering amplitude including the new term generated by the
two-derivative three-body force is shown schematically in Fig.1.
Two amplitudes get mixed: $t_s$ describes the $d_t + N\rightarrow
d_s + N$ process, and $t_t$ describes the $d_t + N\rightarrow d_t +
N$ process, where $d_s$ ($d_t$) is an auxiliary field of two
nucleons in a relative singlet-S (triplet-S) wave.

\begin{eqnarray}
 t_s(p,k)& =  \frac{1}{4}\left[3\mathcal{K}(p,k)
+2\mathcal{H}(E,\Lambda)\right]+\dis\frac{1}{2\pi}
 \int\limits_0^\Lambda \dd q\; q^2
    & \left[\mathcal{D}_s(q)\left[\mathcal{K}(p,q)+2\mathcal{H}(E,\Lambda)
      \right]
t_s(q)\right.\nonumber\\
       &&\left.+\mathcal{D}_t(q)\left[3\mathcal{K}(p,q)+2\mathcal{H}(E,\Lambda)
       \right]
t_t(q)\right] \label{int_equation_triton}\nonumber\\
 t_t(p,k)& = \frac{1}{4}\left[\mathcal{K}(p,k)
+2\mathcal{H}(E,\Lambda)\right]+\dis\frac{1}{2\pi}
 \int\limits_0^\Lambda \dd q\; q^2
  &\left[ \mathcal{D}_t(q)\left[
\mathcal{K}(p,q)+2\mathcal{H}(E,\Lambda)\right]
t_t(q)\right.\nonumber\\
       & & \left.+\mathcal{D}_s(q)
       \left[3\mathcal{K}(p,q)+2\mathcal{H}(E,\Lambda)\right]
t_s(q)\right]\;\;,
\end{eqnarray}
where $\mathcal{D}_{s,t}(q)=\mathcal{D}_{s,t}(E-\frac{q^2}{2M},q)$
are the propagators of the auxiliary fields $d_{s,t}$, and
$\mathcal{K}$ the propagator of the exchanged nucleon, projected
into the S-wave. For the spin-triplet $S$-wave channel, one
determines the two-nucleon interaction up to N$^2$LO by the deuteron
binding momentum $\gamma_t=45.7025\;\mathrm{MeV}$ and effective
range $\rho_t=1.764\;\mathrm{fm}$. Because there is no real bound
state in the spin singlet channel of the two-nucleon system, its
free parameters are better determined by the scattering length
$a_s=1/\gamma_s=-23.714\;\mathrm{fm}$ and the effective range
$r_s=2.73\;\mathrm{fm}$ at zero momentum.

\begin{figure}[!t]
\begin{center}
  \includegraphics*[width=.8\textwidth]{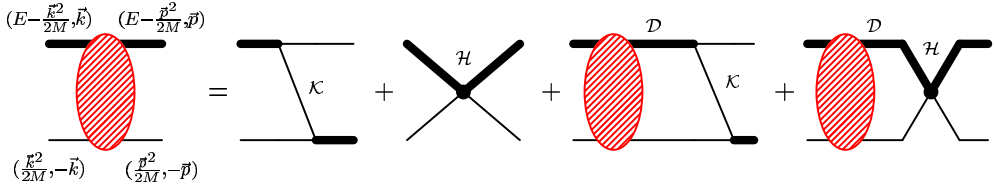}
\caption{The Faddeev equation for $Nd$-scattering. Thick solid
  line is propagator of the two intermediate auxiliary fields $D_s$ and $D_t$,
  denoted by $\calD$; $\calK$:
  propagator of the exchanged nucleon;
  $\calH$: three-body force.}
\label{fig1}
\end{center}
\end{figure}

The neutron-deuteron $J=1/2$ phase shifts $\delta$ is determined
by the on-shell amplitude $t_t(k,k)$, multiplied with the wave
function renormalisation
\begin{equation}
T(k)=Z t_t(k,k)=\frac{3\pi}{M}\frac{1}{k \cot\delta-\ii k}\;\;.
\end{equation}

At thermal energies, the reaction proceeds through $S$-wave capture
predominantly via a magnetic dipole transition, $M^{LSJ}_{i}$, where L=0,
$S$=1/2,3/2 and i=1. To obtain the spin structure, which corresponds to a
definite value of $J$ for the entrance channel, it is necessary to build
special linear combinations of products $\vec DN$ and $\vec \sigma\times\vec
DN$, with $J^{P}=\displaystyle\frac{1}{2}^+$ or
$J^{P}=\displaystyle\frac{3}{2}^+$, and $\vec{D}$ the deuteron spin-one field,
see [15] for details.
$$\vec {\phi}_{1/2}=(i\vec D+\vec\sigma\times\vec D)N~\mbox{and}~(2i\vec
D-\vec\sigma\times\vec D)N\;.$$
 For both possible magnetic dipole transitions with
 $ J^{P}=\displaystyle\frac{1}{2}^+$ (amplitude $g_1$) and
  $J^{P}=\displaystyle\frac{3}{2}^+$ (amplitude $g_3$) we can write:
$$g_1:~~t^\dagger(i\vec D\cdot\vec{e^*}\times\vec k+\vec\sigma\times\vec
D\cdot\vec{e^*}\times\vec k)N,$$
\begin{equation}
g_3:~~t^\dagger(i\vec D\cdot\vec{e^*}\times\vec k+\vec
\sigma\times\vec D\cdot\vec{e^*}\times\vec k)N\;. \label{eq:as8}
\end{equation}

The contribution of the electric transition $E^{LSJ}_i$ for energies of less
than 60 KeV to the total cross section is very small. Therefore, the electric
quadrupole transition $E^{0 (3/2) (3/2)}_2$ from the initial quartet state
will not be considered at thermal energies. The $M_1$ amplitude receives
contributions from the magnetic moments of the nucleon and dibaryon operators
coupling to the magnetic field, which are described by the Lagrange density
\begin{equation}\label{eq:M1}
  \mathcal{L}_B=\frac{e}{2M_N}N^\dagger(k_0+k_1 \tau^3){\sigma.B}
  +e\frac{L_1}{M_N\sqrt{r^{({^1S}_0)}r^{({^3S}_1)}}}{{d_t}^j}^\dagger{{d_s}}_3
  B_j+H.C\;.
\end{equation}
where $k_0=1/2(k_p+k_n)=0.4399$ and $k_1=1/2(k_p-k_n)= 2.35294$ are the
isoscalar and isovector nucleon magnetic moment in nuclear magnetons,
respectively. The NLO-coefficient $L_1$ is fixed at its leading non-vanishing
order to the thermal cross section~\cite{Rupak98}.

\begin{figure}[!htb]
\begin{center}
 \includegraphics*[width=.8\textwidth]{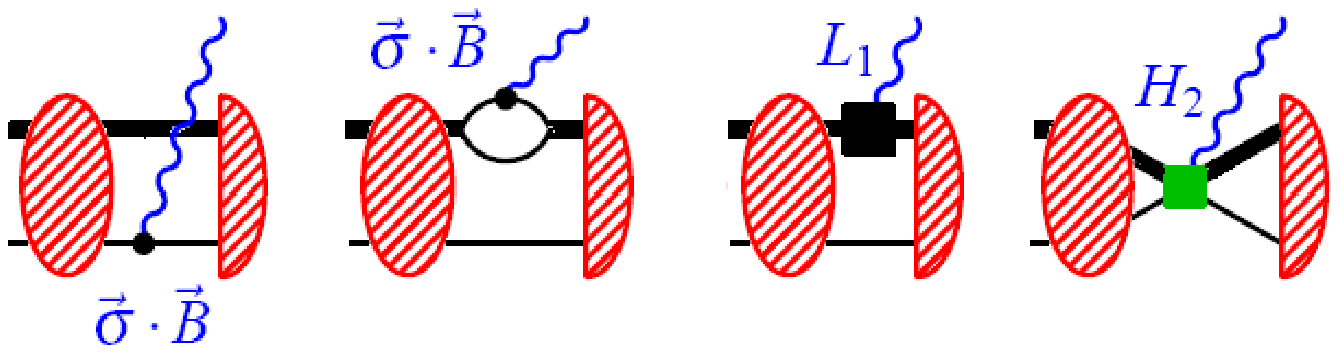}
 \caption{Some diagrams for adding photon-interaction to the Faddeev equation
   up to N$^2$LO.  Wavy line shows photon and small circles show magnetic
   photon interaction. For $L_1$ vertices, see eq.(7); $H_2$:three- body
   force, see eq.(3).  Remaining notation as in Fig.~\ref{fig1}.}. \label{fig2}
\end{center}
\end{figure}

The radiative capture cross section $nd\rightarrow ^3H\gamma$ at
very low energy is given by

\begin{equation}\label{crosssection}
  \sigma=\frac{2}{9}\frac{\alpha}{v_{rel}}\frac{p^3}{4M^2_N}\sum_{iLSJ}
  [{|\widetilde{\chi}^{LSJ}_i|}^2]\;,
\end{equation}
where
\begin{equation}\label{redefine}
  \widetilde{\chi}^{LSJ}_i=\frac{\sqrt{6\pi}}{p\mu_N} \sqrt{4\pi}
  {\chi^{LSJ}_i}\;,
\end{equation}
with $\chi$ stands for either E or M and $\mu_N$ is in nuclear
magneton and p is momentum of the incident neutron in the center
of mass.

We now turn to the Faddeev integral equation to be used in the $M_1$
calculation. We solve the Faddeev equation for nd-scattering and also for the
triton bound state to some order (e.g. LO), then we take these Faddeev
amplitudes and sandwich the photon-interactions with nucleons between them
when the photon kernel is expanded to the same order. This process will be
done separately for NLO and N$^2$LO. Finally the wave function renormalization
in each order will be done.

The diagrams in Fig.~\ref{fig2} represent contributions of electromagnetic
interaction with nucleon, deuteron, four-nucleon-magnetic-photon operator
described by a coupling between the $^3S_1$-dibaryon and $^1S_0$-dibaryon and
a magnetic photon. As mentioned in the introduction, in another
paper~\cite{Sadeghi}, we have presented detailed schematic of these diagrams
in neutron-deuteron radiative capture for ($ 20 \leq E \leq 200 $ keV) up to
N$^2$LO.

The last diagrams in Fig.~\ref{fig2} with insertion of a photon to
the N$^2$LO three-nucleon force $H_2$ vertex is not $M_1$ and we
know that $M_1$ contribution is the dominant contribution at very
low energy and especially for zero energy.  Its contribution should
therefore be very tiny.  Because the leading three-nucleon force
$H_0$ has no derivatives, it is not affected by the minimal
substitution $p \to p - e A$. But the parameter $H_2$ is the
strength of the three-nucleon interaction with two derivatives.
Naturally for the energy range near zero momentum, insertion of
photon to $H_2$ vertices for momentum $p\sim 0.025$eV and $M_1$
transition, could be neglected.  $H_2$ is necessary in
neutron-deuteron scattering to improve cut-off independence but is
defined such that it does not contribute at zero momentum.
Contributions of a photon coupling to $H_2$ are however indeed
negligible at zero energy.

\section{neutron-deuteron radiative capture results at zero energy}
\label{section:results}


We numerically solved the Faddeev integral equation up to N$^2$LO.  We used
$\hbar c=197.327\;\MeV\,\fm$, a nucleon mass of $M=938.918\;\MeV$, for the
$NN$ triplet channel a deuteron binding energy (momentum) of $B=2.225\;\MeV$
($\gamma_d=45.7066\;\MeV$), a residue of $Z_d=1.690(3)$, for the $NN$ singlet
channel an ${}^1\mathrm{S}_0$ scattering length of $a_s=-23.714\;\fm$ ,
$L_1\sim-4.5$ fm by fixing at its leading non-vanishing order by the thermal
cross section.

As in Ref.~[20], we can determine which three-body forces are required at any
given order, and how they depend on the cutoff.

\begin{figure}[!htb]
\begin{center}
  \includegraphics[width=.7\linewidth,clip=true]{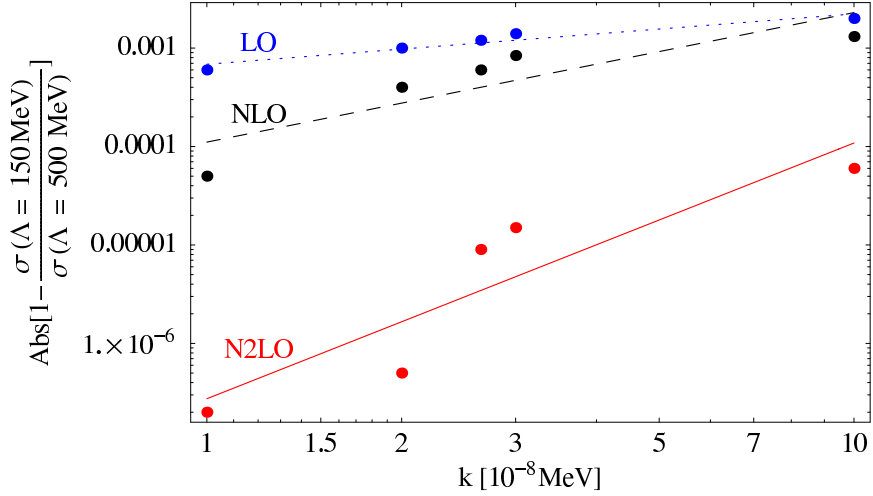}
\end{center}
\vspace*{-0pt} \caption{Curve of  the cutoff variation of  cross
section up to
 N$^2$LO is shown between  $\Lambda=150$ MeV and $\Lambda=500$ MeV. The short dashed, long dashed  and solid line correspond
  to LO, NLO and N$^2$LO , respectively.}
\label{fig4}
\end{figure}

Low-energy observables must be insensitive to the cut-off, namely to
any details of short-distance physics in the region above the
break-down scale of the pion-less EFT, set approximately by the
pion-mass. It was found in Ref.~[20] that no additional
three-nucleon forces are necessary to render a renormalisable
amplitude at N$^2$LO in this process, besides those needed already
in nucleon-deuteron scattering: $H_0$ and $H_2$.  At N$^2$LO , where
we saw that $H_2$ is required, we checked this by varying the
cut-off between $150$ and $500$ MeV. This is a reasonable estimate
of the errors of our calculation due to higher-order effects. As
seen in Fig.~3, in the thermal energy range the cutoff variation is
very small and decreases steadily as we increase the order of the
calculation and it is of the order of $(k/\Lambda)^n,
(\gamma/\Lambda)^n$, where $n$ is the order of the calculation and
$\Lambda=150$ MeV is the smallest cutoff used ( see
Table~\ref{tab:b} and Fig.~\ref{fig4}). Also, errors due to cutoff
variation is decreasing when the order of calculation is increased
up to N$^2$LO.
\begin{table}[!htb]
\caption{ Results for the cutoff variation of  cross section up to
 N$^2$LO is shown between  $\Lambda=150$ MeV and $\Lambda=500$ MeV. }
\label{tab:b} \vspace{0.25cm}
\begin{center}
\begin{tabular}{c c||c|c|c}
\hline & E($10^{-8}$MeV) & LO & NLO & N$^2$LO  \\
 \hline \hline

 & 1    & 0.0006 & 0.00005  & 0.0000002 \\
 & 2    & 0.0010 & 0.00040  & 0.0000005\\
 & 2.65 & 0.0012 & 0.00060  & 0.0000090\\
 & 3    & 0.0014 & 0.00084  & 0.0000150\\
 & 10   & 0.0020 & 0.00131  & 0.0000600\\
 \hline
\end{tabular}
\end{center}
\end{table}

We determined the two-nucleon parameters from the deuteron binding energy,
triplet effective range (defined by an expansion around the deuteron pole, not
at zero momentum), the singlet scattering length, effective range (defined by
expanding at zero momentum), and two body capture process(obtained with
comparison between experimental data and theoretical results for $np
\rightarrow d \gamma$ process at zero energy~\cite{Rupak99}). We fix the
three-body parameters as follows: because we defined $H_2$ such that it does
not contribute at zero momentum scattering, one can first determine $H_0$ from
the ${}^2\mathrm{S}_\frac{1}{2}$ scattering length
$a_3=(0.65\pm0.04)\;\mathrm{fm}$~\cite{doublet_sca}. At LO and NLO, this is
the only three-body force. At N$^2$LO, $H_2$ is required. It is determined by
the triton binding energy $B_3=8.48\;\mathrm{MeV}$. Finally, we solve by
insertion of the potential at a given order in the integral equation and
iteration of kernel.

\begin{figure}[!htb]
\begin{center}
  \includegraphics[width=.6\linewidth,clip=true]{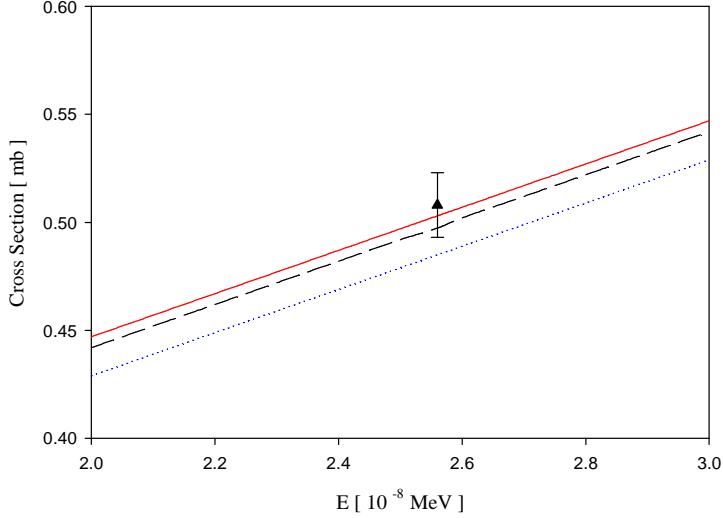}
\end{center}
\vspace*{-0pt} \caption{The cross section for neutron radiative
capture by deuteron as function of the center-of-mass
  kinetic energy $E$ in MeV. The short dashed, long dashed  and solid line correspond
  to the contribution of $M_1$ capture cross section up to LO, NLO and N$^2$LO , respectively.
  Single point shows
 experimental results for this cross section at 0.025 eV~\cite{Jurney}.}
\label{fig3}
\end{figure}

The cross section for neutron-deuteron radiative capture as function
of the center-of-mass energy up to N$^2$LO is shown in
Fig.~\ref{fig3}. We also show single point that shows the available
experimental results for this cross section at 0.025
eV~\cite{Jurney}.

\begin{table}[!htb]
\caption{Comparison between different  theoretical results for
Neutron radiative capture by deuteron at zero energy (0.0253 ev).
Last row shows our EFT result. The last line quotes deviation between data [1] and theory, if it is larger than the theoretical or experimental uncertainty.} \label{tab1} \vspace{0.25cm}
\begin{center}
  \begin{tabular}{c||c|c c}
    Theory  & $\sigma$(mb) & deviation from exp.\\
    \hline \hline
    AV14/VIII (IA+MI+MD) [3] & 0.509&\\
    AV18/IX (IA+MI+MD) [3]   & 0.489&4$\%$\\
    AV14/VIII(IA+MI+MD+$\Delta_{PT}$)~\cite{Viviani} & 0.658 & 29$\%$ \\
    AV18/IX(IA+MI+MD+$\Delta_{PT}$)~\cite{Viviani} & 0.631 & 24$\%$  \\
    AV14/VIII(IA+MI+MD+$\Delta$) ~\cite{Viviani} & 0.600 & 18$\%$ \\
    AV18/IX(IA+MI+MD+$\Delta$)~\cite{Viviani} & 0.578 & 14$\%$  \\
    AV18/IX (gauge inv.) [19] &0.523&3$\%$\\
    AV18/IX (gauge inv. + 3N-current) [19]  &  0.556&\\
    EFT(LO) & 0.485   &  5$\%$ \\
    EFT(NLO) & 0.496   &  \\
    EFT(N$^2$LO) & $0.503\pm0.003$   &   \\
    Experiment~\cite{Jurney}  & $0.508\pm0.015$ \\
    \hline
\end{tabular}
\end{center}
\end{table}

Table~\ref{tab1} shows Comparison between results of different
models-dependent, model-independent EFT and experiment, for neutron radiative
capture by deuteron up to N$^2$LO, at zero energy (0.0253 ev). The
calculations by Viviani et al.~\cite{Viviani,19} shows sensitivity to
short-range physics namely to details of including the physics of the Delta
and pion-exchange currents. The calculation of Ref.~[19] with manifestly
gauge-invariant current operators is quite sensitive to including
meson-exchange three-nucleon currents. One might therefore have been tempted
to conclude that a new three-nucleon force is also needed in the pion-less
EFT.  As shown above, this is not the case: There are no new three-nucleon
forces besides those already fixed in $nd$ scattering at the same order. The
contribution from the photon coupling to a three-nucleon force is negligible
in our calculation. As our result is model-independent and universal, any
model with the same input must -- within the accuracy of our calculation --
lead to the same result. Our inputs are the first two terms of the
effective-range expansion in the singlet- and triplet-S wave of NN scattering,
the proton and neutron magnetic moments, the triton binding energy and nd
scattering length in the doublet-S-wave, and finally the thermal cross section
of the reaction $np\to d\gamma$ (determining $L_1$). More work is needed to
understand why the potential-model calculations [3,19] have the same input but
do not seem to reproduce the same result.

Addressing convergence of the EFT calculation, we notice that the
contributions which are characterized as higher-order in the
power-counting are indeed small: The LO result is $0.485$ mb, with
NLO adding $0.011$ mb, and N$^2$LO another $0.007$ mb. Cut-off
dependence is negligible. The typical size of the expansion
parameter in the pion-less EFT is about $\gamma_t/m_\pi\approx 1/3$.
We therefore estimate the uncertainty from leaving out corrections
at N$^3$LO and higher as about 1/3 of the N$^2$LO correction or
$0.003$ mb.

\section{Conclusion}
\label{section:conclusion}
The cross section for radiative capture of neutrons by deuterons
$nd\rightarrow \gamma{^3H}$ at zero energies with was calculated pionless
Effective Field Theory, the unique, model independent and systematic
low-energy version of QCD for processes involving momenta below the pion mass.
We applied pionless EFT to find numerical results for the $M_1$ contributions.
Incident thermal neutron energies have been considered for this capture
process. At these energy our calculation is dominated by only $S$-wave state
and magnetic transition $M_1$ contribution. The $M_1$ amplitude is calculated
up to Next-to-Next to leading order N$^2$LO. Three-Nucleon forces are needed
up to N$^2$LO order for cut-off independent results. The triton binding energy
and nd scattering length in the triton channel have been used to fix them.
Hence the cross-section is in total determined as
$\sigma_{tot}=[0.485(LO)+0.011(NLO)+0.007(N^2LO)]=[0.503\pm 0.003]mb$. It
converges order by order in low energy expansion. It is also cut-off
independent at this order. We notice that our calculation has a systematic
uncertainty from higher-order terms which is now smaller than the experimental
error-bar.

\section*{Acknowledgment}

We thank L. Marcucci and A. Kievsky for enlightening discussions which spurred
this work.


\bibliographystyle{apsrev}

\end{document}